\documentclass[3p, preprint]{elsarticle}

\journal{Astroparticle Physics}

\usepackage[english]{babel}
\usepackage{units}
\usepackage{graphicx}
\usepackage{amsmath}
\usepackage{amssymb}
\usepackage{caption}
\usepackage{subcaption}

\usepackage{tikz}
\usepackage{pgf}
\usepackage{3dplot} %requires 3dplot.sty to be in same directory, or in your LaTeX installation
\usepackage{caption}
\usepackage{subcaption}

\DeclareMathOperator{\sgn}{sgn}

\definecolor{dark-gray}{gray}{0.3}

\usepackage{subfig}

\usetikzlibrary{shapes,arrows}

\tikzstyle{block} = [rectangle, draw, fill=blue!20,
    text width=5em, text centered, rounded corners, minimum height=4em]
\tikzstyle{line} = [draw, -latex']
\tikzstyle{cloud} = [draw, rectangle,fill=red!20, node distance=3cm,
    minimum height=2em]

\begin{document}

\title{The shape of the radio wavefront of extensive air showers as measured with LOFAR}

\author[ru]{A.~Corstanje\corref{cor1}}
\cortext[cor1]{Principal corresponding author}
\ead{A.Corstanje@astro.ru.nl}
\author[ru]{P.~Schellart\corref{cor2}}
\cortext[cor2]{Corresponding author}
\ead{P.Schellart@astro.ru.nl}
\author[ru,ni]{A.~Nelles\corref{cor2}}
\ead{A.Nelles@astro.ru.nl}
\author[ru]{S.~Buitink}
\author[ru]{J.~E.~Enriquez}
\author[ru,as,ni,mpifr]{H.~Falcke}
\author[as]{W.~Frieswijk}
\author[ru,ni]{J.~R.~H\"orandel}
\author[ru]{M.~Krause}
\author[ru]{J.~P.~Rachen}
\author[rug]{O.~Scholten}
\author[ru]{S.~ter Veen}
\author[ru]{S.~Thoudam}
\author[rug]{G.~Trinh}
\author[ru]{M.~van den Akker}
\author[1]{A.~Alexov}
\author[2]{J.~Anderson}
\author[3,4]{I.~M.~Avruch}
\author[5]{M.~E.~Bell}
\author[6]{M.~J.~Bentum}
\author[7]{G.~Bernardi}
\author[8]{P.~Best}
\author[9]{A.~Bonafede}
\author[2]{F.~Breitling}
\author[10]{J.~Broderick}
\author[9]{M.~Br\"uggen}
\author[11]{H.~R.~Butcher}
\author[12]{B.~Ciardi}
\author[9]{F.~de Gasperin}
\author[6,13]{E.~de Geus}
\author[6]{M.~de Vos}
\author[6]{S.~Duscha}
\author[14]{J.~Eisl\"offel}
\author[15]{D.~Engels}
\author[6]{R.~A.~Fallows}
\author[16]{C.~Ferrari}
\author[6,17]{M.~A.~Garrett}
\author[18,19]{J.~Grie\ss{}meier}
\author[6]{A.~W.~Gunst}
\author[6]{J.~P.~Hamaker}
\author[14]{M.~Hoeft}
\author[20]{A.~Horneffer}
\author[17]{M.~Iacobelli}
\author[21]{E.~Juette}
\author[22]{A. ~Karastergiou}
\author[20]{J.~Kohler}
\author[6,23]{V.~I.~Kondratiev}
\author[20]{M.~Kuniyoshi}
\author[6]{G.~Kuper}
\author[6]{P.~Maat}
\author[2]{G.~Mann}
\author[6]{R. McFadden}
\author[24,25]{D.~McKay-Bukowski}
\author[6,4]{M.~Mevius}
\author[6]{H.~Munk}
\author[6]{M.~J.~Norden}
\author[6]{E.~Orru}
\author[26]{H.~Paas}
\author[27]{M.~Pandey-Pommier}
\author[6]{V.~N.~Pandey}
\author[6]{R.~Pizzo}
\author[6]{A.~G.~Polatidis}
\author[20]{W.~Reich}
\author[17]{H.~R\"ottgering}
\author[10]{A.~M.~M.~Scaife}
\author[28]{D.~Schwarz}
\author[29,30]{O.~Smirnov}
\author[22]{A.~Stewart}
\author[2]{M.~Steinmetz}
\author[31]{J.~Swinbank}
\author[18]{M.~Tagger}
\author[6]{Y.~Tang}
\author[32]{C.~Tasse}
\author[6]{C.~Toribio}
\author[6]{R.~Vermeulen}
\author[2]{C. Vocks}
\author[7]{R.~J.~van Weeren}
\author[6]{S.~J.~Wijnholds}
\author[20]{O.~Wucknitz}
\author[6]{S.~Yatawatta}
\author[32]{P.~Zarka}
 
\address[ru]{Department of Astrophysics/IMAPP, Radboud University Nijmegen, P.O. Box 9010, 6500 GL Nijmegen, The Netherlands}
\address[as]{Netherlands Institute for Radio Astronomy (ASTRON), Postbus 2, 7990 AA Dwingeloo, The Netherlands}
\address[ni]{Nikhef, Science Park Amsterdam, 1098 XG Amsterdam, The Netherlands}
\address[mpifr]{Max-Planck-Institut f\"{u}r Radioastronomie, Auf dem H\"ugel 69, 53121 Bonn, Germany}
\address[rug]{University of Groningen, P.O. Box 72, 9700 AB Groningen, The Netherlands}

\address[1]{Space Telescope Science Institute, 3700 San Martin Drive, Baltimore, MD 21218, USA }

\address[2]{Leibniz-Institut f\"{u}r Astrophysik Potsdam (AIP), An der Sternwarte 16, 14482 Potsdam, Germany }

\address[3]{SRON Netherlands Insitute for Space Research, PO Box 800, 9700 AV Groningen, The Netherlands }

\address[4]{Kapteyn Astronomical Institute, PO Box 800, 9700 AV Groningen, The Netherlands }

\address[5]{ARC Centre of Excellence for All-sky astrophysics (CAASTRO), Sydney Institute of Astronomy, University of Sydney Australia }

\address[6]{Netherlands Institute for Radio Astronomy (ASTRON), Postbus 2, 7990 AA Dwingeloo, The Netherlands }

\address[7]{Harvard-Smithsonian Center for Astrophysics, 60 Garden Street, Cambridge, MA 02138, USA }

\address[8]{Institute for Astronomy, University of Edinburgh, Royal Observatory of Edinburgh, Blackford Hill, Edinburgh EH9 3HJ, UK }

\address[9]{University of Hamburg, Gojenbergsweg 112, 21029 Hamburg, Germany }

\address[10]{School of Physics and Astronomy, University of Southampton, Southampton, SO17 1BJ, UK }

\address[11]{Research School of Astronomy and Astrophysics, Australian National University, Mt Stromlo Obs., via Cotter Road, Weston, A.C.T. 2611, Australia }

\address[12]{Max Planck Institute for Astrophysics, Karl Schwarzschild Str. 1, 85741 Garching, Germany }

\address[13]{SmarterVision BV, Oostersingel 5, 9401 JX Assen }

\address[14]{Th\"{u}ringer Landessternwarte, Sternwarte 5, D-07778 Tautenburg, Germany }

\address[15]{Hamburger Sternwarte, Gojenbergsweg 112, D-21029 Hamburg }

\address[16]{Laboratoire Lagrange, UMR7293, Universit\`{e} de Nice Sophia-Antipolis, CNRS, Observatoire de la C\'{o}te d'Azur, 06300 Nice, France }

\address[17]{Leiden Observatory, Leiden University, PO Box 9513, 2300 RA Leiden, The Netherlands }

\address[18]{LPC2E - Universite d'Orleans/CNRS }

\address[19]{Station de Radioastronomie de Nancay, Observatoire de Paris - CNRS/INSU, USR 704 - Univ. Orleans, OSUC , route de Souesmes, 18330 Nancay, France }

\address[20]{Max-Planck-Institut f\"{u}r Radioastronomie, Auf dem H\"ugel 69, 53121 Bonn, Germany }

\address[21]{Astronomisches Institut der Ruhr-Universit\"{a}t Bochum, Universitaetsstrasse 150, 44780 Bochum, Germany }

\address[22]{Astrophysics, University of Oxford, Denys Wilkinson Building, Keble Road, Oxford OX1 3RH }

\address[23]{Astro Space Center of the Lebedev Physical Institute, Profsoyuznaya str. 84/32, Moscow 117997, Russia }

\address[24]{Sodankyl\"{a} Geophysical Observatory, University of Oulu, T\"{a}htel\"{a}ntie 62, 99600 Sodankyl\"{a}, Finland }

\address[25]{STFC Rutherford Appleton Laboratory,  Harwell Science and Innovation Campus,  Didcot  OX11 0QX, UK }

\address[26]{Center for Information Technology (CIT), University of Groningen, The Netherlands }

\address[27]{Centre de Recherche Astrophysique de Lyon, Observatoire de Lyon, 9 av Charles Andr\'{e}, 69561 Saint Genis Laval Cedex, France }

\address[28]{Fakult\"{a}t fr Physik, Universit\"{a}t Bielefeld, Postfach 100131, D-33501, Bielefeld, Germany }

\address[29]{Department of Physics and Elelctronics, Rhodes University, PO Box 94, Grahamstown 6140, South Africa }

\address[30]{SKA South Africa, 3rd Floor, The Park, Park Road, Pinelands, 7405, South Africa }

\address[31]{Astronomical Institute 'Anton Pannekoek', University of Amsterdam, Postbus 94249, 1090 GE Amsterdam, The Netherlands }

\address[32]{LESIA, UMR CNRS 8109, Observatoire de Paris, 92195   Meudon, France }

\begin{abstract}
Extensive air showers, induced by high energy cosmic rays impinging on the Earth's atmosphere, produce radio emission that is measured with the LOFAR radio telescope.
As the emission comes from a finite distance of a few kilometers, the incident wavefront is non-planar. A spherical, conical or hyperbolic shape of the wavefront has been proposed, but measurements of individual air showers have been inconclusive so far. 
For a selected high-quality sample of 161 measured extensive air showers, we have reconstructed the wavefront by measuring pulse arrival times to sub-nanosecond precision in 200 to 350 individual antennas.
For each measured air shower, we have fitted a conical, spherical, and hyperboloid shape to the arrival times. The fit quality and a likelihood analysis show that a hyperboloid is the best parametrization.
Using a non-planar wavefront shape gives an improved angular resolution, when reconstructing the shower arrival direction.
Furthermore, a dependence of the wavefront shape on the shower geometry can be seen. This suggests that it will be possible to use a wavefront shape analysis to get an additional handle on the atmospheric depth of the shower maximum, which is sensitive to the mass of the primary particle.
\end{abstract}

\maketitle

\section{Introduction}
A high-energy cosmic ray that enters the atmosphere of the Earth will interact with a nucleus of an atmospheric molecule. This interaction produces secondary particles, which in turn interact, thereby creating a cascade of particles: an \emph{extensive air shower}. The origin of these cosmic rays and their mass composition are not fully known.

Due to the high incident energy of the cosmic ray, the bulk of the secondary particles propagate downward with a high gamma factor. As this air shower passes through the atmosphere and the Earth's magnetic field, it emits radiation, which can be measured by antennas on the ground in a broad range of radio frequencies  (MHz - GHz)  \cite{Allan1966,Jelley1965,Falcke2005}. For a review of recent developments in the field see \cite{Huege:2013a}. The measured radiation is the result of several emission processes \cite{CoREAS}, and is further influenced by the propagation of the radiation in the atmosphere with non-unity index of refraction \cite{EVA}. Dominant in the frequency range considered in this study is the interaction in the geomagnetic field \cite{Kahn1966,Allan:1971,Falcke2005,Codalema2009}. An overview of the current understanding of the detailed emission mechanisms can be found in \cite{Huege2012}.

The radio signal reaches the ground as a coherent broadband pulse, with a duration on the order of 10 to $\unit[100]{ns}$ (depending on the position in the air shower geometry). As the radio emission originates effectively from a few kilometers in altitude, the incident wavefront as measured on the ground is non-planar. Geometrical considerations suggest that the amount of curvature and the shape of the wavefront depend on the height of the emission region, suggesting a relation to the depth of shower maximum, $X_{\rm max}$.
The depth of shower maximum is related to the primary particle type.

Assuming a point source would result in a spherical wavefront shape, which is used for analysis of LOPES data \cite{Nigl2008}. It is argued in \cite{Schroeder2011} that the actual shape of the wavefront is not spherical, but rather conical, as the emission is not point-like but stretched along the shower axis. In a recent further refinement of this study, based on CoREAS simulations, evidence is found for a hyperbolic wavefront shape (spherical near the shower axis, and conical further out) \cite{LOPESwavefront:2014}. Hints for this shape are also found in the air shower dataset collected by the LOPES experiment \cite{Lopes2012}. However, due to high ambient noise levels, the timing precision of these measurements did not allow for a distinction between spherical, hyperbolical and conical shapes on a shower-by-shower basis, and only statistically was a hyperbolic wavefront shape favored.

\begin{figure}
\begin{center}
\includegraphics[width=0.80\textwidth]{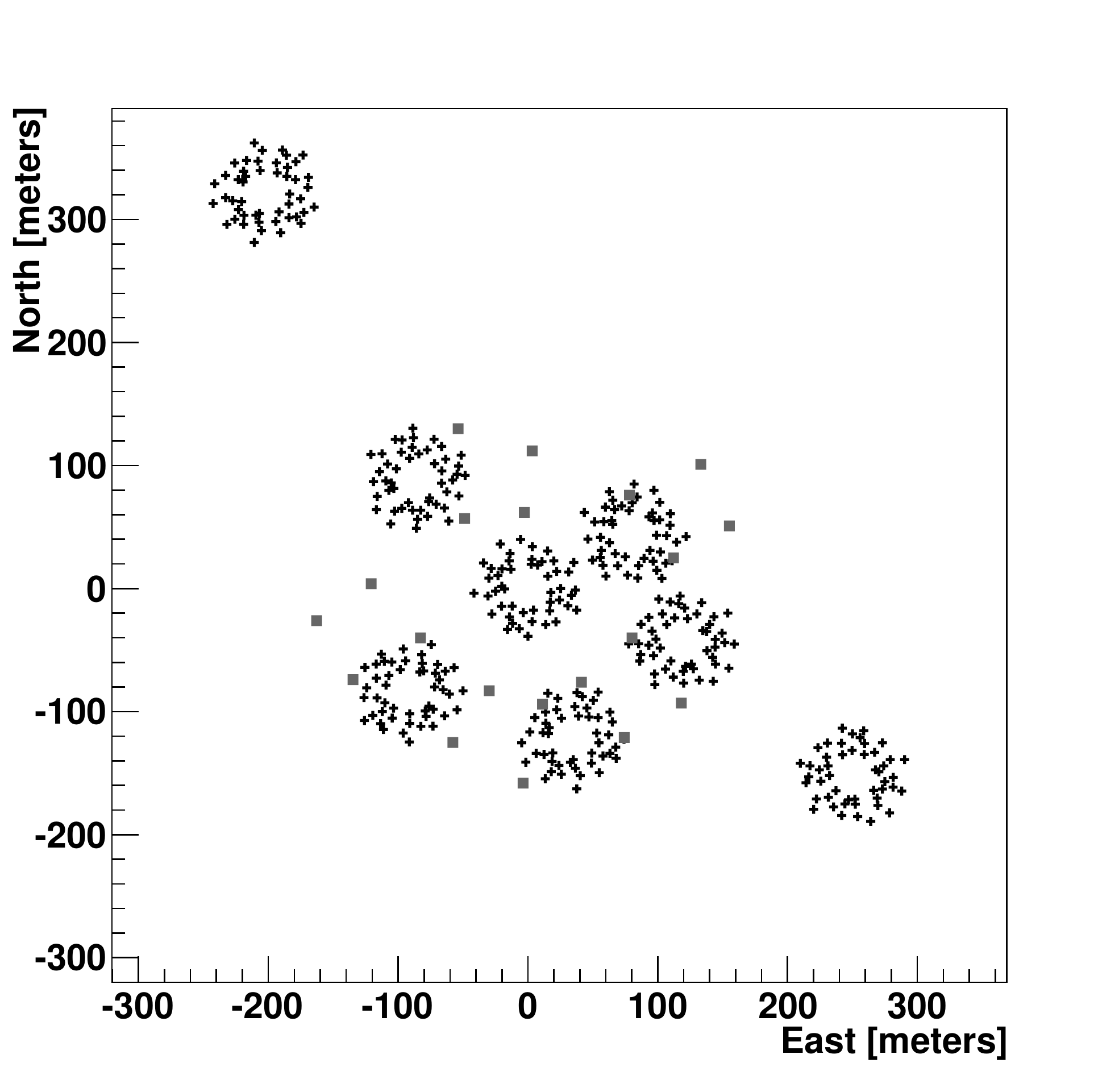}
\caption{Layout of the innermost 8 stations of LOFAR. For each station the outer ring of low band radio antennas (black plus symbols), used for the analysis in this paper, are depicted. Located with the innermost six stations are the particle detectors (grey squares) used to trigger on extensive air showers.}
\label{fig:core_layout}
\end{center}
\end{figure}

We use the LOFAR radio telescope \cite{LOFAR} to measure radio emission from air showers, in order to measure wavefront shapes for individual showers.
LOFAR consists of an array of two types of antennas: the low-band antennas (LBA)  sensitive to frequencies in a bandwidth of $\unit[10-90]{MHz}$, and the high-band antennas (HBA) operating in the $\unit[110-240]{MHz}$ range. While air showers have been measured in both frequency ranges \cite{Schellart:2013,Nelles:2014}, this study only uses data gathered with the $\unit[10-90]{MHz}$ low-band antennas. A combination of analog and digital filters limits the effective bandwidth to $\unit[30-80]{MHz}$ which has the least amount of radio frequency interference.
For detecting cosmic rays we use the (most densely instrumented) inner region of LOFAR, the layout of which is depicted in Fig.\ \ref{fig:core_layout}.
LOFAR is equipped with ring buffers (called Transient Buffer Boards) that can store the raw-voltage signals of each antenna for up to 5~seconds. These are used for cosmic-ray observations as described in \cite{Schellart:2013}.

Inside the inner core of LOFAR, which is a circular area of $\unit[320]{m}$ diameter, an array of 20 scintillator detectors (LORA) has been set up \cite{Thoudam:2014}.
This air shower array is used to trigger a read-out of the Transient Buffer Boards at the moment an air shower is detected. The buffer boards provide a raw voltage time series for every antenna in a LOFAR station (a group of typically 96 LBA plus 48 HBA antennas that are processed together in interferometric measurements), in which we identify and analyze the radio pulse from an air shower.
Analysis of the particle detections delivers basic air shower parameters such as the estimated position of the shower axis, energy, and arrival direction.

The high density of antennas of LOFAR, together with a high timing resolution ($\unit[200]{MHz}$ sampling rate) are especially favorable for measuring the wavefront shape.

\section{Simplified model for the wavefront shape}

\begin{figure}
	\begin{center}
	\begin{subfigure}[b]{0.5\textwidth}
	\includegraphics[width=\textwidth]{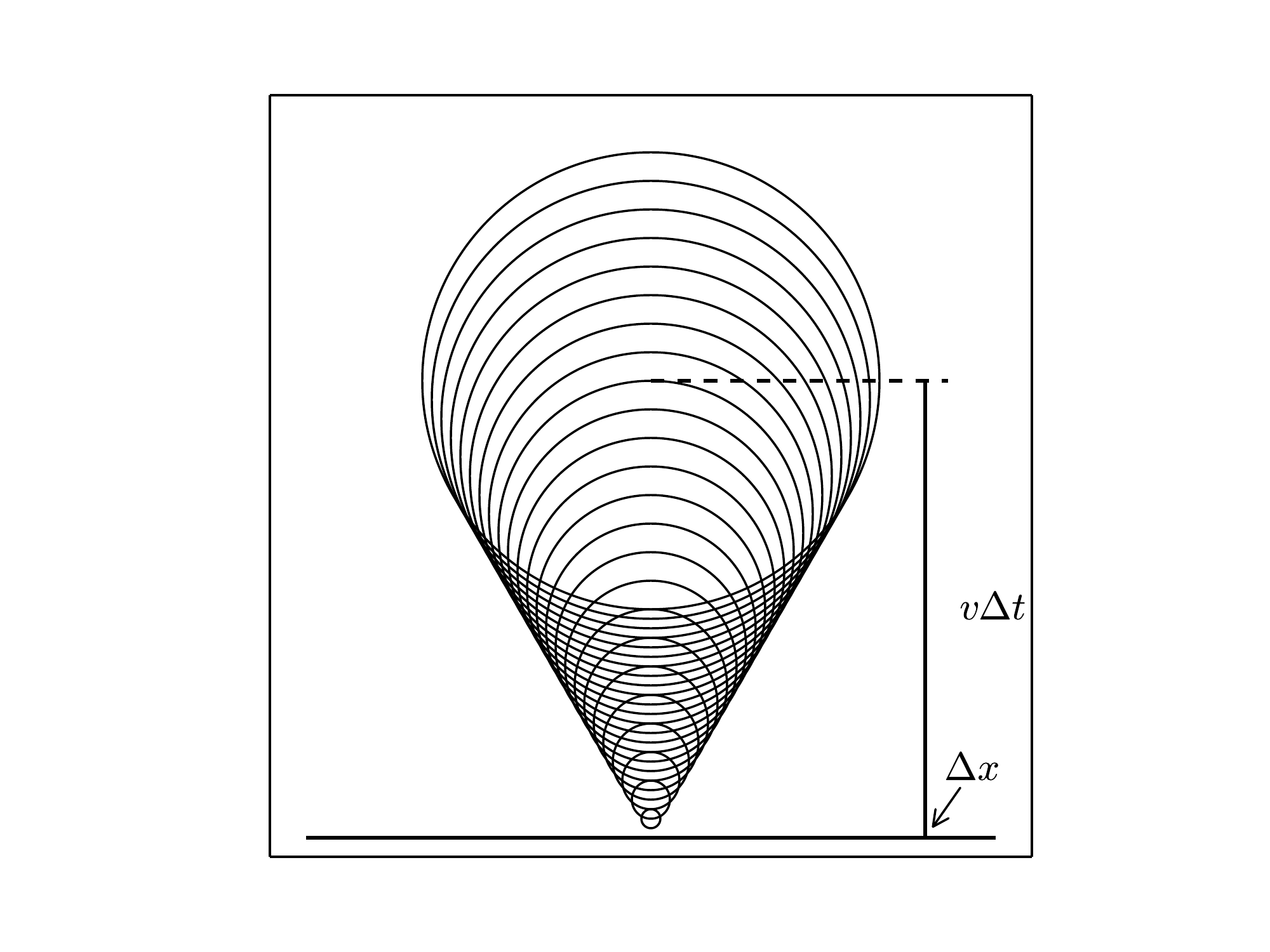}
	\subcaption{Small}
	\label{fig:why_near}
	\end{subfigure}
	\begin{subfigure}[b]{0.5\textwidth}
	\includegraphics[width=\textwidth]{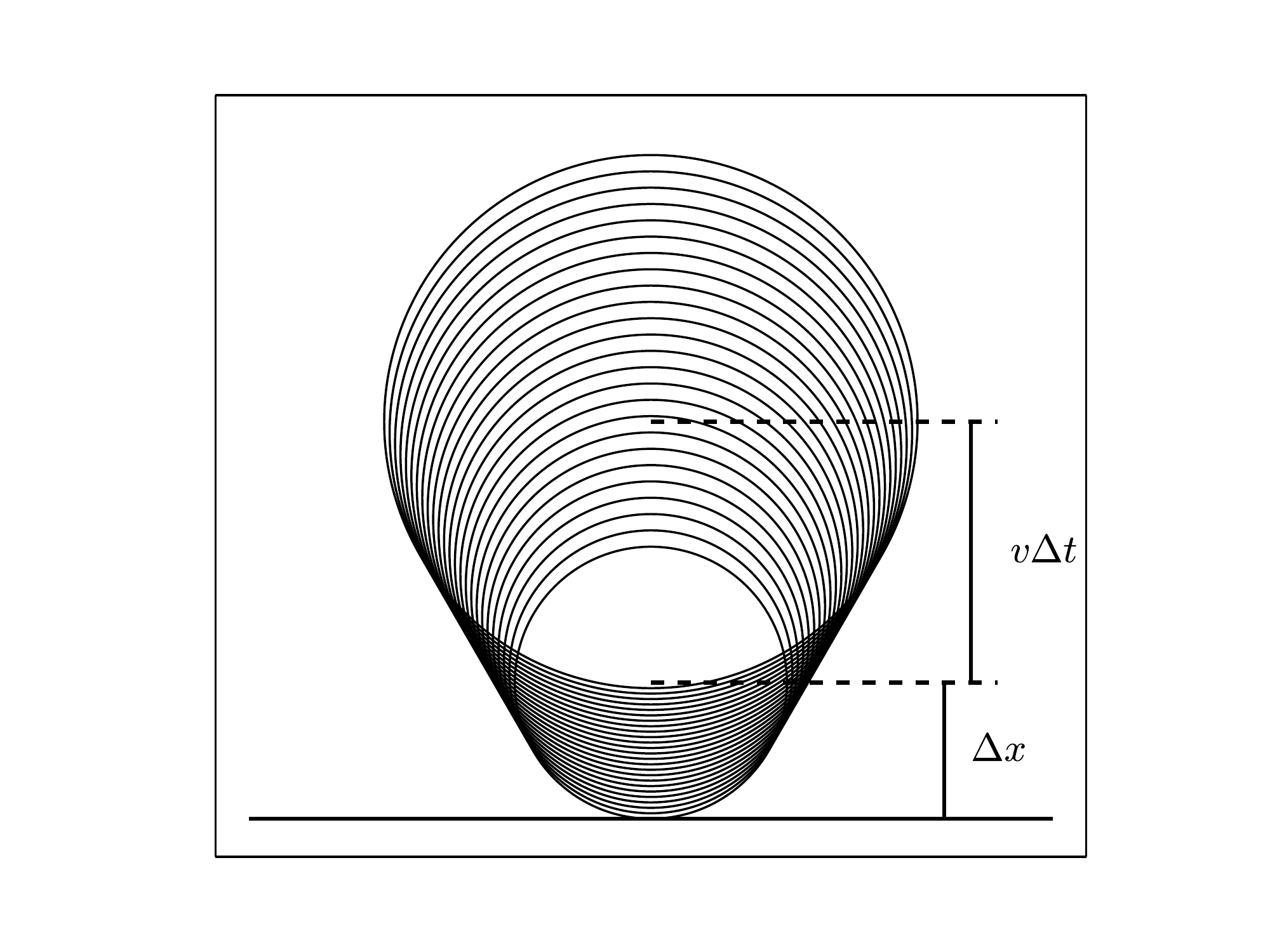}
	\subcaption{Intermediate}
	\label{fig:why_intermediate}
	\end{subfigure}
	\begin{subfigure}[b]{0.5\textwidth}
	\includegraphics[width=\textwidth]{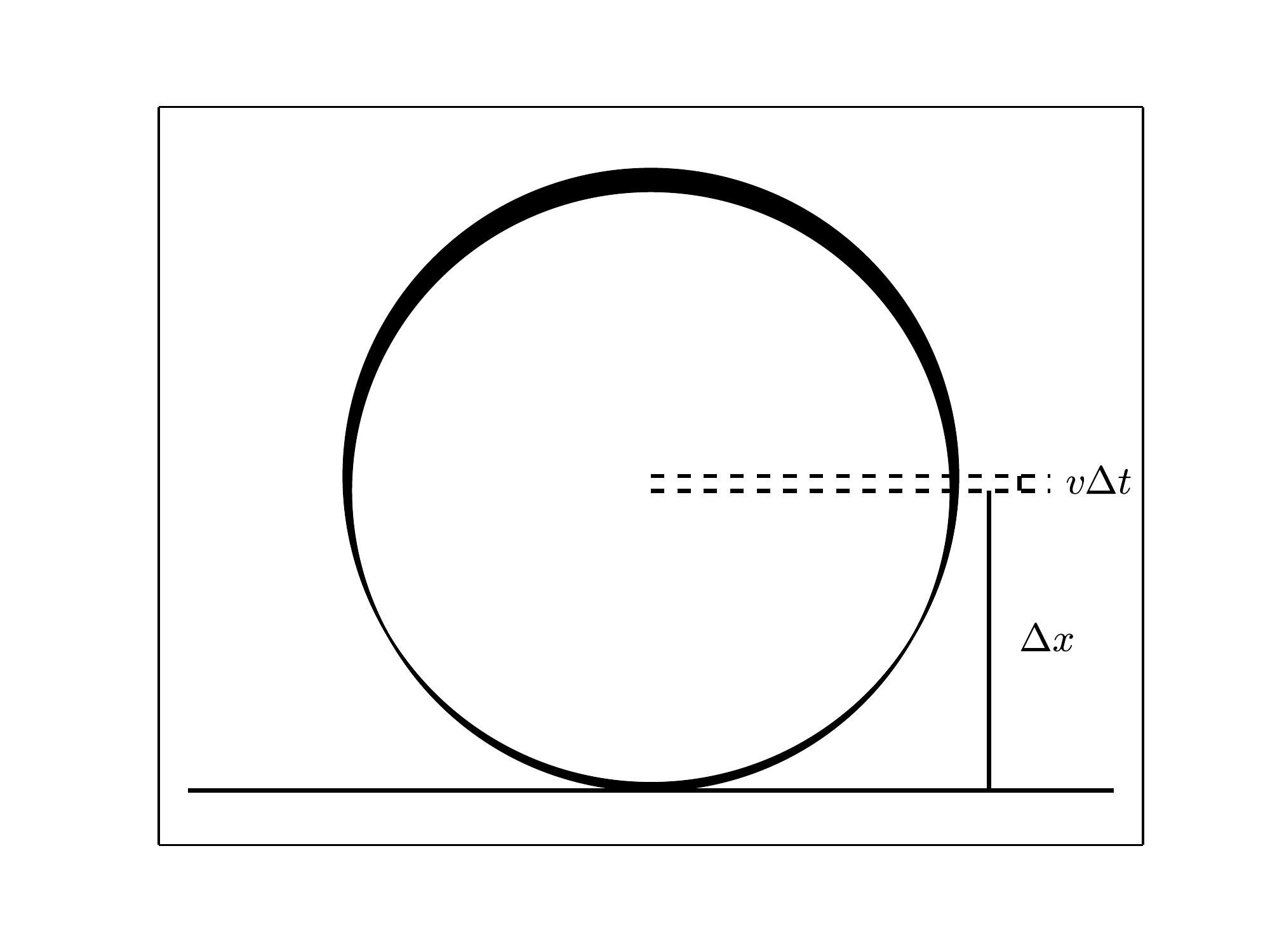}
	\subcaption{Large}
	\label{fig:why_far}
	\end{subfigure}
	\end{center}
	\caption{Toy model motivating a hyperbolic wavefront shape. A point source moves vertically at a velocity $v > c/n$ and emits for a limited amount of time. The solid horizontal line represents the ground plane. The generated wavefront is observed as conical (top panel) by an observer at small distances to the point where the source stops emitting. Observers at intermediate distances see a hyperbolic wavefront shape (middle panel). For observers at larger distances the observed wavefront shape is closer to a sphere (bottom panel).}
\label{fig:why_hyperbolic}
\end{figure}

Inspection of the pulse arrival times in our datasets (as explained in the following sections) shows that while the shape at larger distances from the shower axis might be described by a conical wavefront, in many measured air showers there is significant curvature near the shower axis. A natural choice for a function of two parameters that describes this behavior is a hyperbola. In Fig.\ \ref{fig:why_hyperbolic} a toy model is sketched, where the wavefront is formed by assuming the emission to be generated by a point source. At any given time, during the emission generation, the source generates a spherical wavefront that expands at the local speed of light $c/n$, where $n$ is the local index of refraction. The point source is moving at a velocity $v > c/n$, and emits radiation for a limited amount of time $\Delta t$. In real extensive air showers this corresponds to the duration in which the bulk of the radiation is generated. The radiation is measured by an observer at a distance $\Delta x$ from the point where the emission stopped. When this distance is small, $\Delta x / (v\, \Delta t) < 1$, (top panel) the combined wavefront shape is approximately conical. Even so, unless the distance to the last emission point is zero, a small curvature is visible near the shower axis, the radius of curvature corresponding to the distance. When viewed at intermediate distances, $\Delta x / (v\, \Delta t) \approx 1$, (middle panel), the opening angle of the conical part increases and the curved part near the shower axis extends a bit further outward. This shape is closely approximated by a hyperbola.
Only when the distance to the last emission point is very large compared to the duration of the emission times the local speed of light, $\Delta x / (v\, \Delta t) \gg 1$, is the wavefront shape approximately spherical. In this simplified picture, with constant but non-unity index of refraction, the wavefront shape is thus hyperbolic for most observer distances. We expect the general characteristics of this simplified model to hold even for a realistic atmosphere where the refractive index changes with height as the main criterion $n>1$ still holds true.

Motivated by this toy model we therefore compare three parametrizations of the wavefront shape: a sphere, a cone and a hyperboloid, and evaluate the quality of the fits to the LOFAR measurements.

\section{Measurements}
For this analysis we have used air-shower measurements with LOFAR accumulated between June 2011 and November 2013. These consist of 2 to 5 milliseconds of raw voltage time series for every antenna of the LOFAR core stations; we identify the air shower's radio pulse in every individual trace, and measure its strength and arrival time.

In order to have a dense, high-quality sampling of the radio wavefront, and a substantial distance range of more than $\sim\unit[150]{m}$, we require an air shower to be detected in at least four LOFAR core stations. Furthermore, the highest quality data are obtained with the outer ring of low-band antennas, as its more sparse layout gives a more even distribution of measurements over the area of interest.
Therefore, the sample is restricted to this subset. This leaves a total of 165 measured air showers. Of these 165, three fail calibration of time differences between stations due to corrupted data (see Sect.\ \ref{sec:station_time_calibration}) and one is unreliable due to thunderstorm conditions (see Sect.\ \ref{sec:thunderstorm}). This leaves a total of 161 high quality air shower measurements for this analysis.

All measured air showers are processed by the standard cosmic-ray reconstruction software as described in \cite{Schellart:2013}.
\subsection{Pulse arrival times \& uncertainties}
\label{sec:measurements}
The arrival time of the radio pulse in each dipole is determined using the raw-voltage traces. We define the arrival time as the time of the pulse maximum in the amplitude (or Hilbert) envelope of the analytic signal $A(t)$: 
\begin{equation}
 A(t) = \sqrt{x^2(t) + \hat{x}^2(t)},
\end{equation}
where $\hat{x}(t)$ is the Hilbert transform of the voltage-trace signal $x(t)$, upsampled by a factor of 32. The Hilbert transform is defined by 
\begin{equation}
\mathcal{F}\left[\hat{x}(t)\right](\omega) = -i\; \sgn (\omega)\; \mathcal{F}\left[x(t)\right](\omega),
\end{equation}
where $\mathcal{F}$ is the Fourier transform. This allows for arrival time measurements at a much higher time resolution than suggested by the $\unit[200]{MHz}$ sampling rate ($\unit[5]{ns}$ sampling period).

The attainable timing precision varies with the signal-to-noise ratio ($S/N$).
Uncertainties in the arrival time are assigned independently to each datapoint using the measured $S/N$ in amplitude following
\begin{equation}
\sigma_{t_{\mathrm{max}}} = \frac{12.65}{S/N}\;\rm{ns}.
\label{eq:uncertainty_relation}
\end{equation}
A similar relation was found in \cite{Schroder:2012}. The one used here is derived from the data for each antenna, using a procedure as follows.

\begin{figure}
\begin{center}
\includegraphics[width=0.80\textwidth]{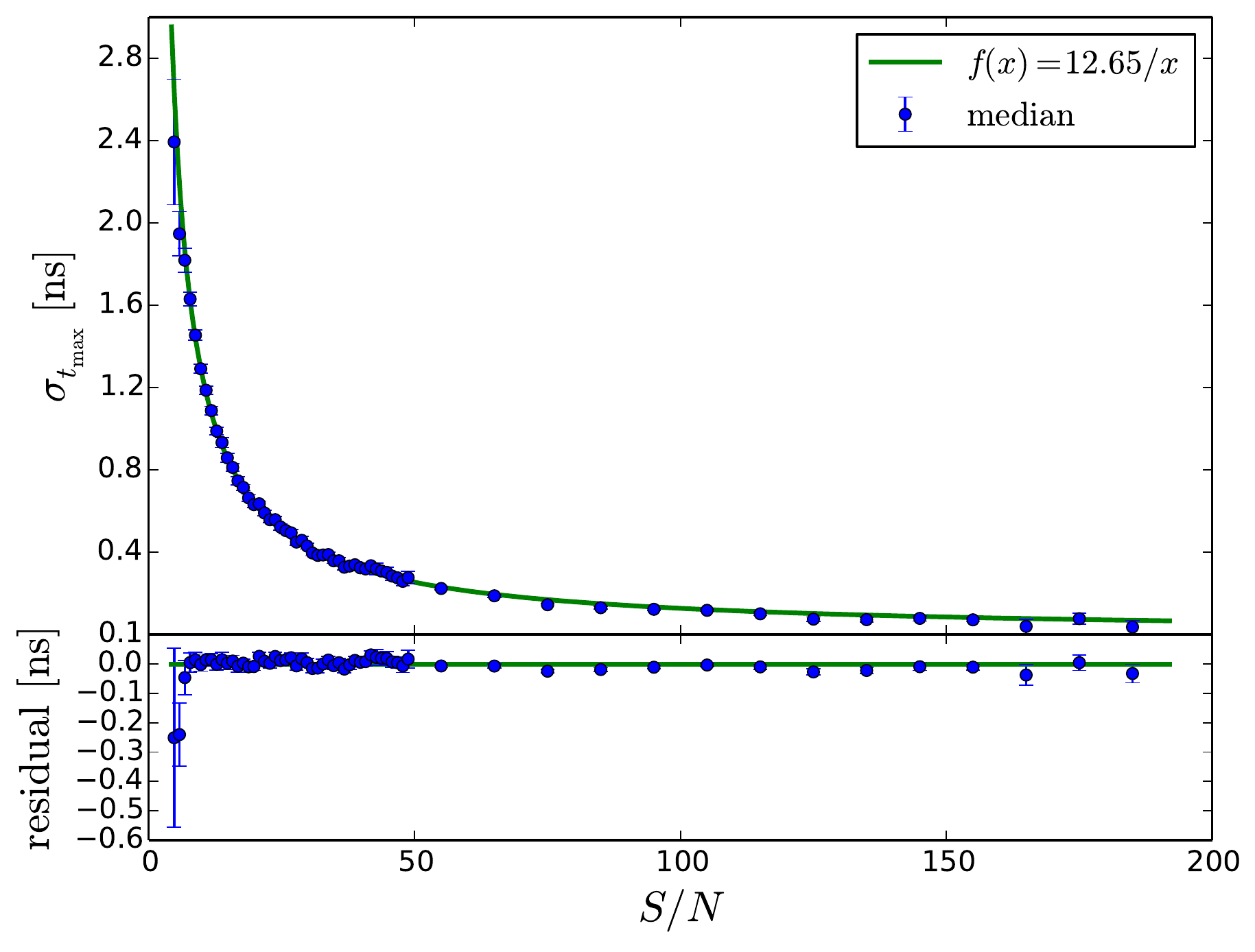}
\caption{Uncertainties on the determination of the pulse arrival time as a function of signal to noise of the pulse amplitude. For each $S/N$ bin (of width $1.0$ for $S/N < 50$ and width $10.0$ for $S/N \geq 50$) the circular dots (blue) give the median value of the uncertainties. Error bars represent the standard error on the median in each bin. The solid line (green) represents the fitted relation; the lower panel shows the residual to the fit.}
\label{fig:snr_timing}
\end{center}
\end{figure}

Uncertainties on the timing arise from distortions of the pulse shape due to fluctuations in the background noise. These uncertainties on the timing can be determined from the amplitude data for each antenna. To quantify them, we first select a noise block outside of the pulse region for each antenna and calculate the root-mean-square noise level $N$. We also calculate the signal-to-noise ratio $S/N$ of the pulse maximum (in amplitude).
Subsequently we add this noise to the data containing the pulse, and the pulse arrival time is calculated using the procedure described above. This procedure is repeated 10 times, where at each time the noise block is shifted by $\unit[100]{ns}$. This gives 10 measurements of the pulse arrival time. The standard deviation of this set for each antenna is a measure of the uncertainty on the determination of the pulse arrival time. However, because this procedure effectively reduces $S/N$ by a factor of $\sqrt{2}$, pulses with a low $S/N$ are no longer correctly identified.
Therefore, instead of assigning this uncertainty to the datapoint directly, we estimate the uncertainties as a function of $S/N$ by processing data for all antennas in the full air shower dataset. Uncertainty data points are binned with respect to $S/N$. The bin size is set to $1.0$ below $S/N=50$ and to $10.0$ above to ensure a sufficient number of points per bin.
To prevent outliers (due to accidental spike selection in some antennas) from heavily influencing the results, the median and uncertainty on the median are calculated for any given $S/N$ bin.
The result can be seen in Fig.\ \ref{fig:snr_timing}.

One can see that the timing uncertainty is inversely proportional to the signal-to-noise ratio. Fitting this relation gives the proportionality factor in Eq.\  \ref{eq:uncertainty_relation}, and we use this to assign an uncertainty to the arrival time of the pulse maximum for each antenna depending on its measured S/N.

\subsection{Time differences between stations}
\label{sec:station_time_calibration}
For time calibration of individual antennas within one LOFAR station, we use standard LOFAR calibration tables as described in \cite{Schellart:2013}. Since all LOFAR core stations share a single clock these calibration solutions are stable over time. However, before October 10th 2012 this common clock was only available for the innermost region (consisting of 6 stations). Every other core station had its own clock, synchronized by GPS\footnote{Global Positioning System}. Drifts of these clocks with respect to each other were on the order of $\unit[10]{ns}$ which is large compared to the other timing uncertainties. Datasets taken before this date therefore require a more involved calibration procedure described below.

Every air shower raw voltage trace is only $\unit[2]{ms}$ long, and all measured air showers are scattered over a 2-year timespan. Therefore, using dedicated calibration observations is not feasible, as these are typically planned in advance and can take minutes to hours.
Instead, we make use of radio transmitter signals present in every dataset. These transmitters emit continuous waves, measured at each antenna with a different phase. We use the Fast Fourier Transform to calculate the phase per antenna and frequency channel. The phase differences between antenna pairs can be used (directly) to monitor and correct for deviations from a trusted timing calibration; this technique was originally developed by the LOPES experiment as demonstrated in \cite{Schroeder:2010}.

In addition to this, we make use of the (known) position of the transmitter in order to predict the relative phases at every antenna pair (accounting for the geometric delays). 
The difference in measured versus expected (relative) phase, averaged over each LOFAR station, yields the inter-station clock calibration.
It was found that higher-frequency signals can be measured with better phase stability; the public radio transmitters at $88$ to $\unit[94]{MHz}$ are suitable for this purpose. 

We use the (strongest) $\unit[88.0]{MHz}$ transmitter to fix the station's clock offsets modulo its period, $\unit[11.364]{ns}$. The remaining ambiguity is resolved using trial and error in the wavefront analysis, incrementally adding outer-core stations to the (already calibrated) inner core data. As $\unit[11]{ns}$ is large compared to the expected wavefront arrival time delays between stations, there is only one best-fitting solution.

Differences in filter characteristics or propagation effects between antennas in a station are expected to average out over an entire station, leading to a calibration of the station clock offsets to about $\unit[0.3]{ns}$ precision.

\subsection{Shower parameters}
An independent reconstruction of the shower is performed based on the detected particle density in each scintillator detector, as described in \cite{Thoudam:2014}. This yields the direction and location of the shower axis, as well as an energy estimate. However, these reconstructed values are only reliable for a restricted parameter space; for example the shower axis should fall inside the instrumented area. In order to retain a substantial set of showers for this analysis, these cuts are not applied to the set of radio measurements. Therefore, we do not have reliably reconstructed shower axis locations for all measured air showers and the core positions are only used as initial estimates that are optimized later (see Sect.~\ref{sec:fit_stability}). The reconstructed direction inferred from particle densities is however independent of the quality of other reconstructed quantities.

\section{Reconstructing the wavefront shape}
From the arrival time of the pulse in different radio antennas, and the information from the particle detector array, we find the shape of the wavefront using the following procedure.

\subsection{Plane-wave approximation and curvature}
We infer the general direction of the incoming pulse by obtaining the best-fitting plane wave solution to the arrival times of the radio pulse:
\begin{equation}
c\, t_i = A x_i + B y_i + C,
\end{equation}
where $t_i$ and $(x_i, y_i)$ are the arrival times and antenna positions respectively. This holds for an antenna array for which all antennas lie in the same plane at constant $z$, which is true for LOFAR's inner core region.
The fitted parameters $A$ and $B$ yield the azimuth angle $\phi$ and zenith angle $\theta$, from the relations
\begin{align}
\theta & = \arcsin \left(\sqrt{A^2 + B^2}\right), \\
\phi & = \frac{\pi}{2} - \arctan(B / A),
\end{align}
where the angle from the arctangent is taken in the appropriate quadrant.
The global offset $C$ is not used here.
We can subtract the arrival times of the best-fitting plane wave from the measured times.
This gives the curvature of the wavefront with respect to the array barycenter, defined as the average of the $(x,y)$ coordinates for antennas with data.

\subsection{Shower plane geometry}
\label{sec:shower_plane_geometry}
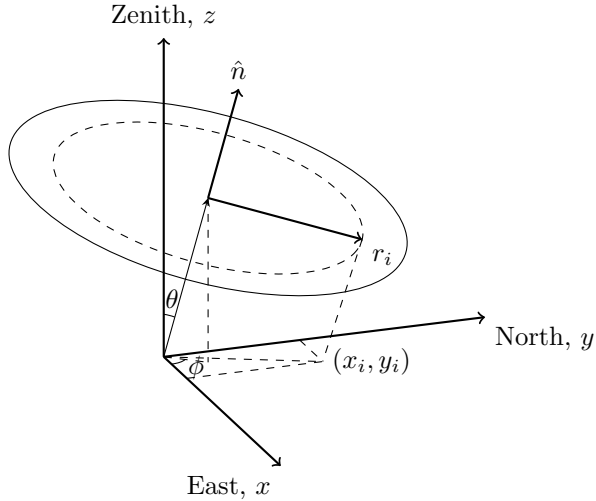
\begin{figure}
\centering

% only change this block to change projection the rest is calculated automatically
\tdplotsetmaincoords{70}{70}
\pgfmathsetmacro{\rvec}{0.8}
\pgfmathsetmacro{\thetavec}{15}
\pgfmathsetmacro{\phivec}{50}
\pgfmathsetmacro{\radiusveclen}{.7}
\pgfmathsetmacro{\radiusvecangle}{15}

\begin{tikzpicture}[scale=3.0,tdplot_main_coords]
\coordinate (O) at (0,0,0);
\tdplotsetcoord{P}{\rvec}{\thetavec}{\phivec};
\tdplotsetcoord{L}{cos(\thetavec * pi / 180) * \radiusveclen * \radiusveclen / sqrt((\radiusveclen * cos((\phivec + \radiusvecangle) * pi / 180))^2 + (cos(\thetavec * pi / 180) * \radiusveclen * sin((\phivec + \radiusvecangle) * pi / 180))^2)}{90}{(\phivec + \radiusvecangle)};
\draw[dashed] (O) -- (L);
\draw[thick,->] (0,0,0) -- (1.5,0,0) node[anchor=north east]{East, $x$};
\draw[thick,->] (0,0,0) -- (0,1.5,0) node[anchor=north west]{North, $y$};
\draw[thick,->] (0,0,0) -- (0,0,1.5) node[anchor=south]{Zenith, $z$};
\draw[-stealth] (O) -- (P) node[anchor=south west]{};
\draw[dashed] (O) -- (Pxy);
\draw[dashed] (P) -- (Pxy);
\draw[dashed] (Lx) -- (Lxy);
\draw[dashed] (Ly) -- (Lxy) node[anchor=west]{$(x_i,y_i)$};
\tdplotdrawarc{(O)}{0.1}{0}{\phivec}{}{}
\node at (25:.2){$\phi$};
\tdplotsetthetaplanecoords{\phivec}
\tdplotdrawarc[tdplot_rotated_coords]{(0,0,0)}{0.2}{0}{\thetavec}{anchor = north}{}
\node[tdplot_rotated_coords] at (8:0.28){$\theta$};
\tdplotsetrotatedcoords{\phivec}{\thetavec}{0}
\tdplotsetrotatedcoordsorigin{(P)}
\tdplotdrawarc[tdplot_rotated_coords]{(0,0,0)}{0.9}{0}{360}{}{}
\draw[thick,tdplot_rotated_coords,->] (\radiusvecangle:0) -- (\radiusvecangle:\radiusveclen) node[anchor=north west]{$r_i$};
\draw[dashed,tdplot_rotated_coords] (\radiusvecangle:\radiusveclen) -- (Lxy);
\tdplotdrawarc[tdplot_rotated_coords, dashed]{(0,0,0)}{\radiusveclen}{0}{360}{}{}
\draw[thick,tdplot_rotated_coords,->] (0,0,0) -- (0,0,.55) node[anchor=south]{$\hat{n}$};
\end{tikzpicture}
\caption[Shower plane geometry]{Geometry of the shower plane for a shower arriving from azimuth $\phi$ and zenith angle $\theta$. The direction of the shower plane is defined by its normal vector $\hat{n}$. All antenna positions $(x_i, y_i)$ are projected onto this plane giving $r_i$, the distance to the shower axis.}
\label{fig:shower_plane}
\end{figure}

Given the shower axis position and direction, we can make a one-dimensional plot of the wavefront as a function of the distance to the shower axis. This assumes axial symmetry of the wavefront. In order to do this, all antenna positions are projected into the shower plane (defined by the shower axis as its normal vector), see Fig.\ \ref{fig:shower_plane}.
A one-dimensional function describing the wavefront curvature can then be fitted to the arrival times as a function of distance to the shower axis.

The projection into the shower plane is (by its definition) performed along lines parallel to the shower axis. This is an approximation, as the true wavefront is not planar, but has a small deviation angle $\alpha$ with respect to the shower plane. The angle may depend on the distance to the shower axis but can generally, for large distances be taken close to $\unit[1]{^\circ}$. Projecting perpendicular to the true wavefront rather than the shower plane would give, to first order in $\alpha$, a correction to the projected distance to the shower axis $r$, of $\Delta r / r = \tan(\alpha) \tan(\theta)$. For zenith angles below $\unit[45]{^\circ}$, and for the longest distances in our dataset of about $\unit[500]{m}$, this could introduce a timing uncertainty (scatter) of at most $\unit[0.5]{ns}$ at the largest distances (or $\unit[0.3]{ns}$ for typical zenith angles around 30 degrees). This is comparable to the calibration uncertainty between LOFAR station's clocks (see Sect.~\ref{sec:station_time_calibration}).
A possible bias from an asymmetric antenna layout would be no larger than $2 \%$ on the estimated angle $\alpha$ and would not change the best-fitted shape.

Therefore, for the purpose of this analysis we apply the shower-plane projection, noting that in a detailed comparison with simulations, and for very large and/or inclined air showers, a two-dimensional fitting procedure may be favored.

\subsection{Fitting the wavefront shape}
\label{sec:fitting}
Various wavefront shapes have been proposed; we test a conical and spherical shape, such as argued for in \cite{Schroeder2011}. 
We also test a hyperboloid; this is a natural function with 2 parameters that combines a curved shape for small distances, and a conical shape for large distances.

The fit functions, for the arrival time differences with respect to a plane wave as a function of distance to the shower axis, are  those for a line (cone), a circle (sphere) and a hyperbola (hyperboloid) respectively:
\begin{align}
\label{fit_equations}
c\, t_{\mathrm{con}}(r) & = s\; r, \\
c\, t_{\mathrm{sph}}(r) & = \sqrt{R^2 + r^2} - R, \\
c\, t_{\mathrm{hyp}}(r) & = -a + \sqrt{a^2 + b^2\, r^2}, 
\end{align}
where $s$ is the cone slope, $R$ the radius of the sphere, and $a$ and $b$ are the parameters of the hyperboloid.
These three functions are fitted in a standard non-linear least squares approach; the shower core $x$ and $y$ positions, needed to get the distance $r$, are used as free parameters in the fit, as explained further in the next section.
We keep for each fit type the best fitting parameters as well as the fit quality, as measured by the unreduced $\chi^2$ value
\begin{equation}
\chi_{\mathrm{type}}^2 = \sum_{i=0}^{N_\mathrm{antennas}} \frac{(t_{\mathrm{type}}(r_{i}) - t_{i})^2}{\sigma_{i}^2},
\end{equation}
where $t_i$ is the arrival time of the pulse maximum at antenna $i$ corrected for the best fitting plane wave solution and $\sigma_{i}$ the corresponding uncertainty calculated using equation \ref{eq:uncertainty_relation}.

\subsection{Considerations for fit stability}
\label{sec:fit_stability}
As the arrival time differences from a plane wave solution, and thus the shape of the wavefront, are sensitive to the direction and location of the shower axis, we include these as free parameters in the fitting procedure. If the core position would be well known, e.g.~from signal amplitudes and/or comparison with simulations, each fit would have fewer degrees of freedom. Therefore typically, comparing the fit qualities of each shape, we find a lower bound to the differences with respect to the best fit.

To prevent the fit from becoming unstable or finding only local minima we choose a nested approach. For every trial of the shower axis location, we optimize the direction; for every trial of the direction, we calculate the best-fit curve parameters using a nonlinear least-squares solver.
Furthermore, to prevent the shower axis location search from getting stuck in a local minimum it is first optimized on a $\unit[500]{m}$ by $\unit[500]{m}$ grid in steps of $\unit[100]{m}$ and only in later iterations optimized further using a Nelder-Mead simplex optimization, starting from the optimal grid position.

\subsection{Including particle detector information}
When optimizing the shower axis location, it might happen that the position is not well constrained due to the geometric distribution of the measurements. Furthermore, fitting a non-correct wavefront may also lead to an unphysical shower axis location. Typically this takes the form of the shower axis location moving too far away from the measured barycenter. The data from the particle detectors provide a further constraint on the shower axis location. The lateral distribution of the signal (number of particles as a function of distance to the shower axis) is well described by a Nishimura-Kamata-Greisen function (NKG) \cite{Kamata:1958,Greisen:1960} and will restrict the position of the shower axis.
Therefore, in the fit procedure we also re-fit the particle detector data using an NKG-function and add the (unreduced) $\chi^2$ of this fit to the total $\chi^2$ to minimize. This has a small influence when the shower axis location is within reasonable distances (due to the much larger number of radio measurements) but starts to dominate when the shower axis location moves to a position not supported by particle data. Note that the stored $\chi^2_{\mathrm{type}}$ of the optimal curve fit does \emph{not} include the particle fit $\chi^2$.

\subsection{Thunderstorm observations}
\label{sec:thunderstorm}
It has been reported that measured radio signals of air shower are amplified during thunderstorm conditions, which is attributed to the acceleration of electrons and positron in the electric fields \cite{Buitink2007}. In order to avoid a bias in the wavefront analysis, we have excluded one air shower that was measured during thunderstorm conditions. As the current dataset was taken without a local electric-field meter, the definition of thunderstorm conditions is based on lightning detections as provided by the Royal Dutch Meteorological Institute \cite{Wessels1998}. This cut does not account for local electric fields that do not manifest in lightning strikes. A local electric field meter is currently being installed at LOFAR. Future analyses will investigate the effects of the local electric field on air shower radio emission in more detail.

\section{Results}
\begin{figure}
	\centering
	\includegraphics[trim=1.5cm 0cm 3cm 0cm, clip=true, width=0.80\textwidth]{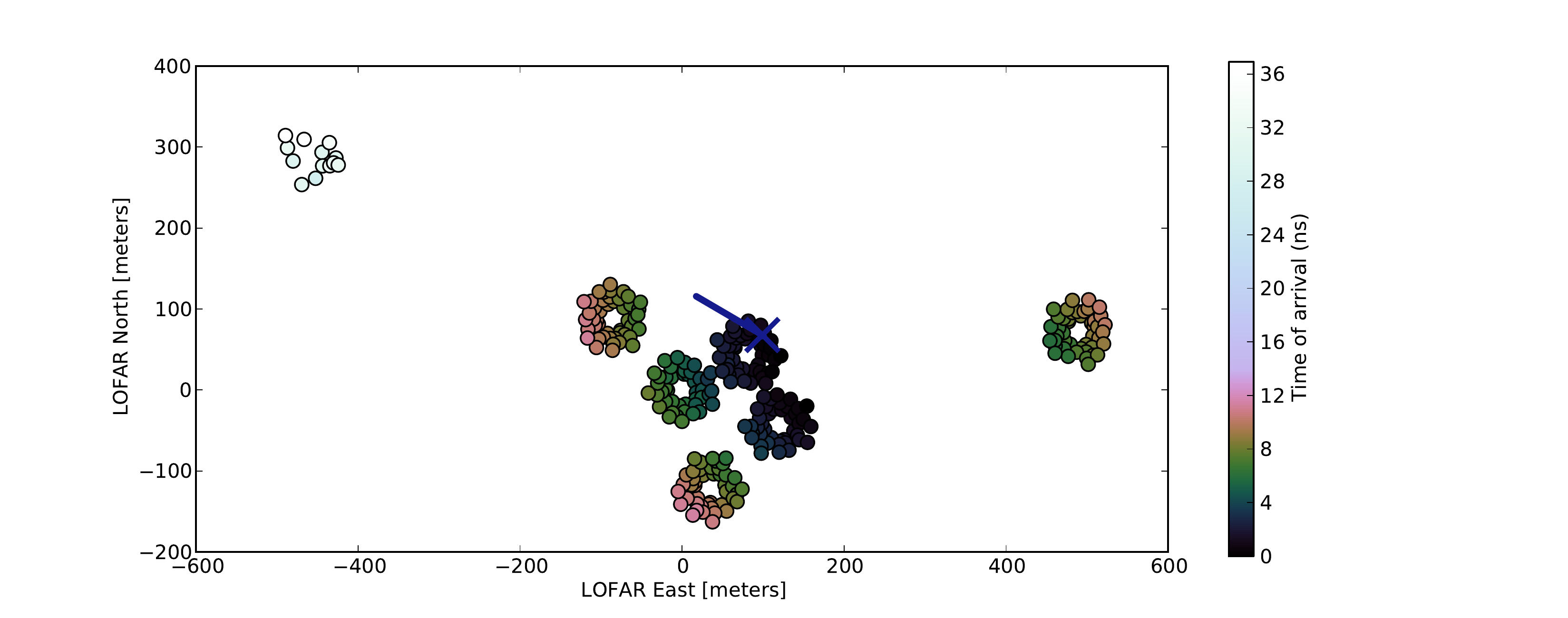}
	\caption[Superterp]{Relative arrival times for an example air shower measured with the LOFAR low band antennas. Circles indicate LBA antenna positions and their color corresponds to the measured pulse delay with respect to the best fitting plane wave solution. The shower axis (as determined from the particle detector data) is indicated by the blue line corresponding to the azimuthal arrival direction and cross where it intersects the ground.}
\label{fig:footprint_curvature}
\end{figure}

An example shower is shown in Fig.\ \ref{fig:footprint_curvature}. This plot shows the layout of the LOFAR low-band antennas in the inner-core region. 
The colors show deviations from the best-fitting plane-wave solution, increasing outward from the center of the array.

\subsection{Wavefront shape}
\begin{figure}
	\centering
	\begin{subfigure}[b]{0.50\textwidth}
	\includegraphics[width=\textwidth]{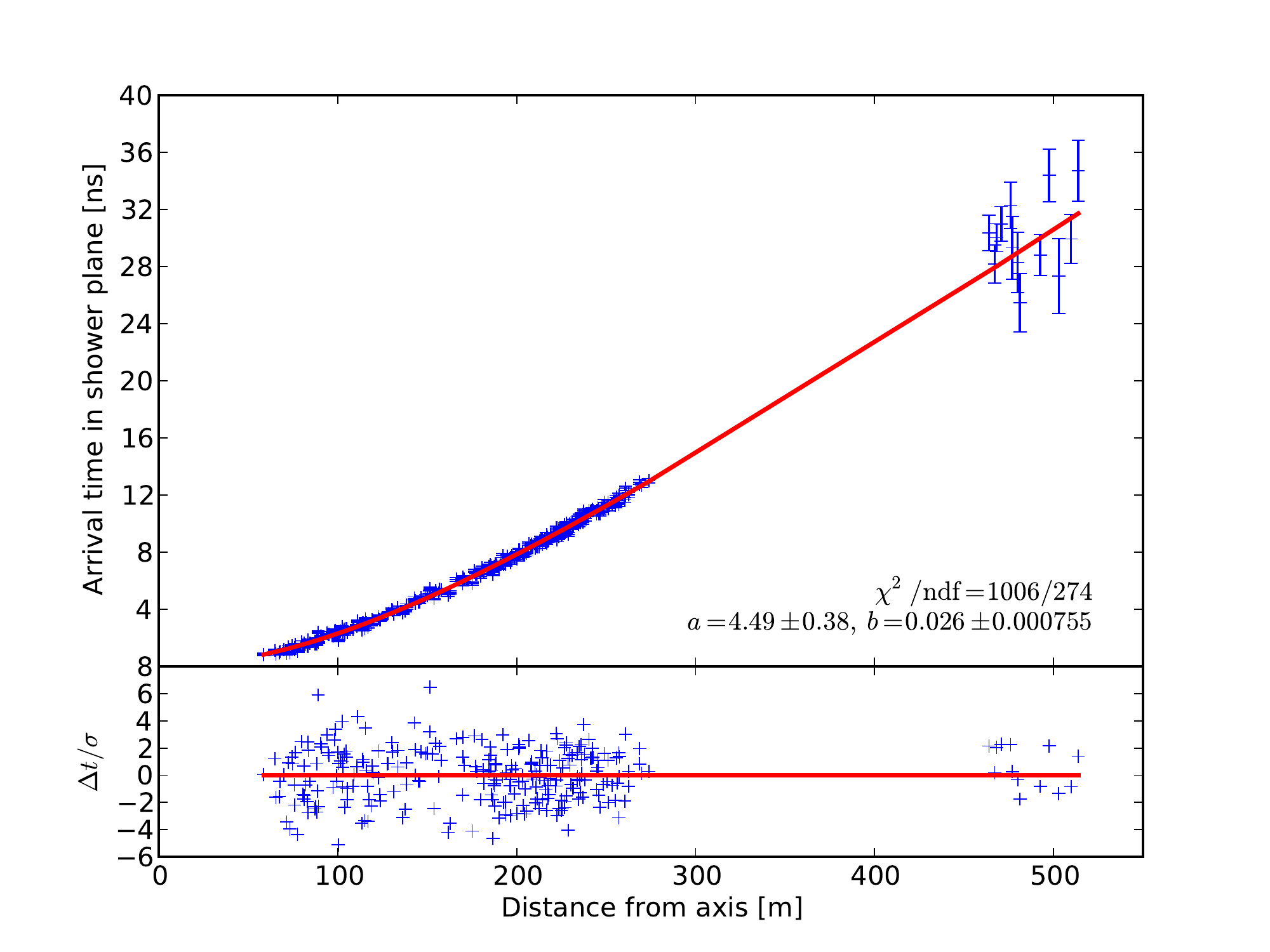}
	\subcaption{Hyperbolic fit}
	\label{fig:hyp_fit}
	\end{subfigure}
	\begin{subfigure}[b]{0.50\textwidth}
	\includegraphics[width=\textwidth]{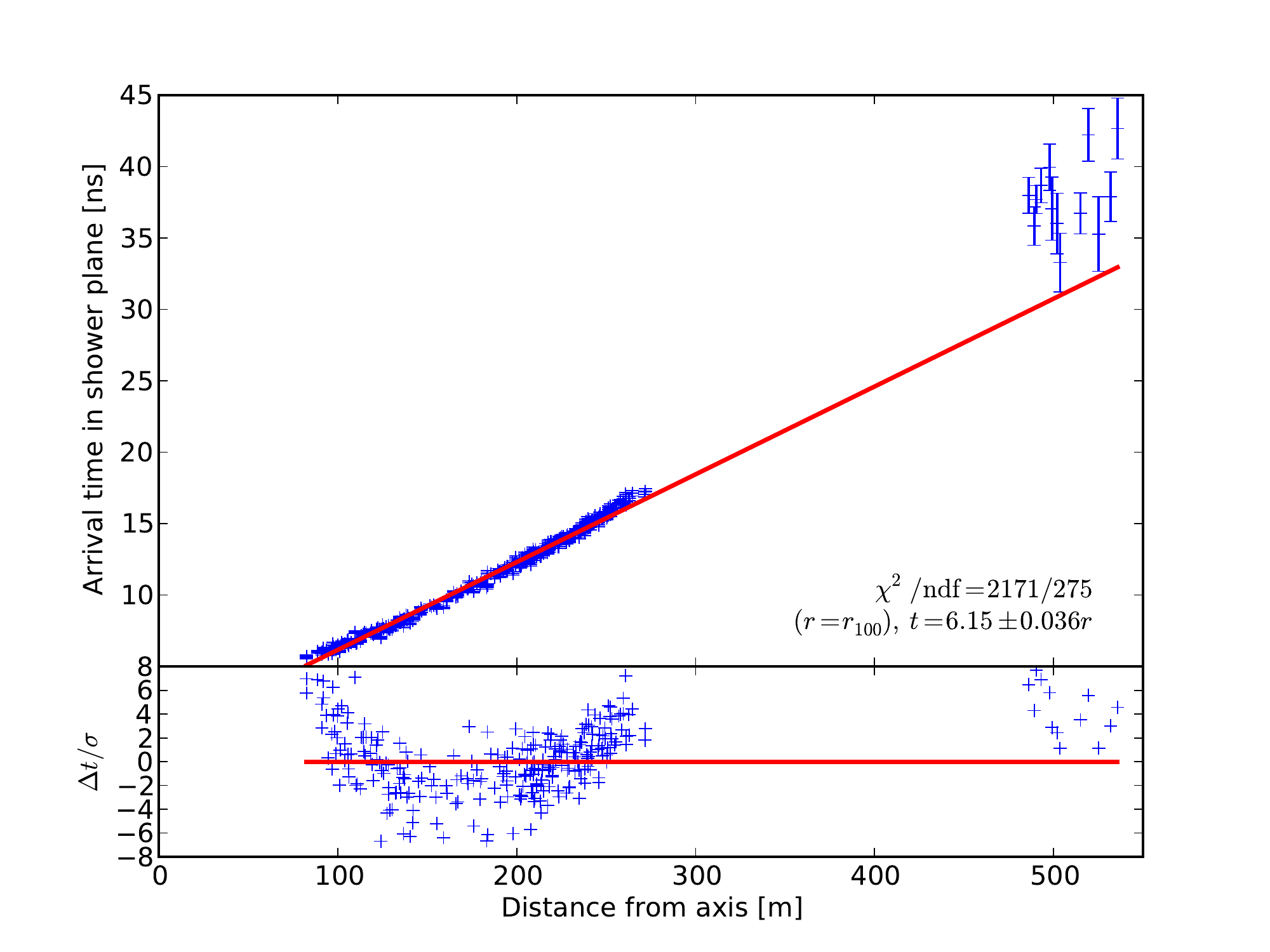}
	\subcaption{Conical fit}
	\label{fig:con_fit}
	\end{subfigure}
	\begin{subfigure}[b]{0.50\textwidth}
	\includegraphics[width=\textwidth]{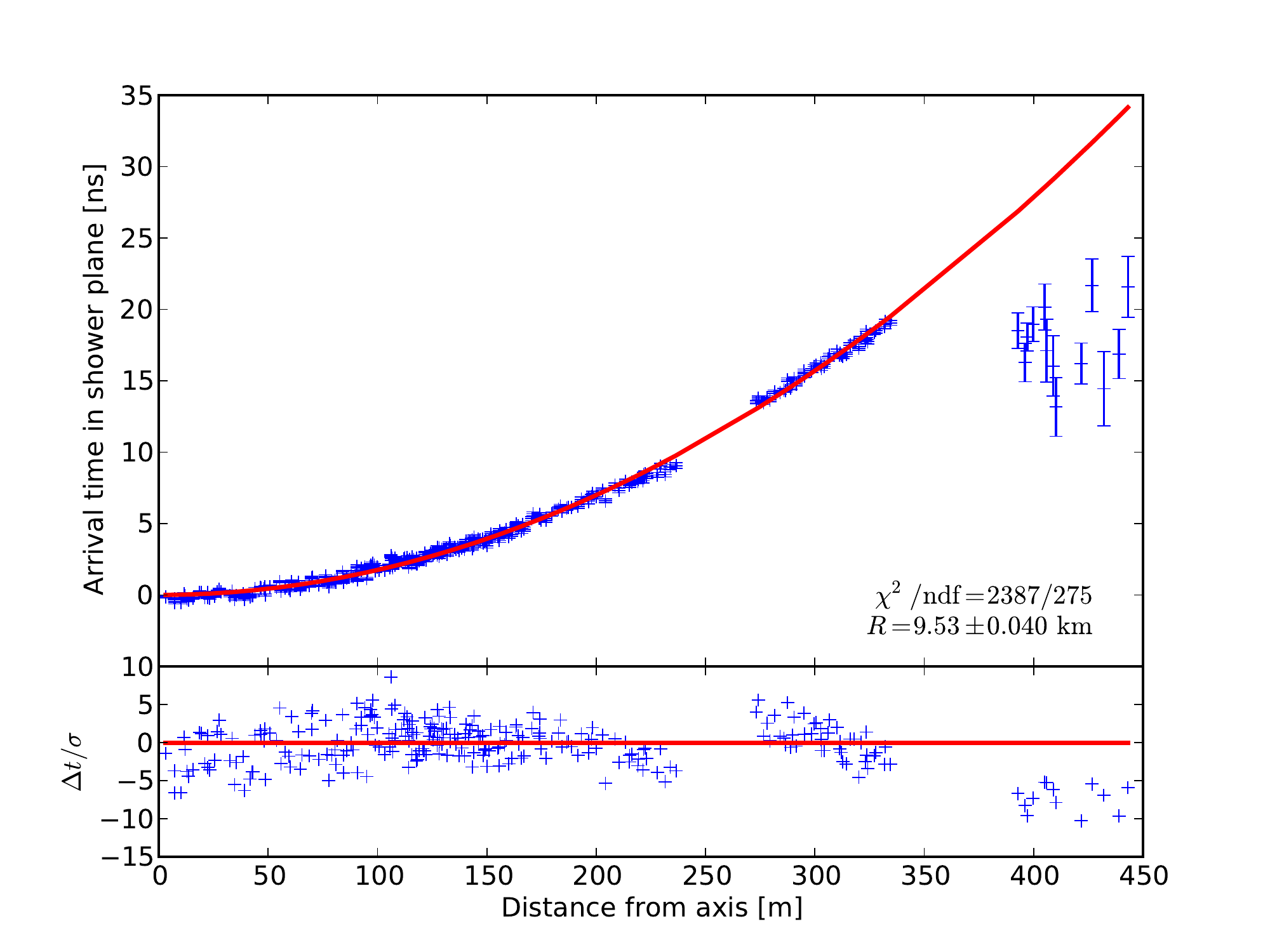}
	\subcaption{Spherical fit}
	\label{fig:sph_fit}
	\end{subfigure}

	\caption[Shape]{The arrival time differences from a plane wave as a function of distance to the shower axis with the best fitting shape solutions. A hyperbolic (top), conical (middle) and spherical (bottom) fit has been applied, respectively. Each plot shows the arrival times as a function of the distance to the shower axis (top panel) and deviations from the best fit scaled to the uncertainty for each datapoint (bottom panel). Note that the shower core position is a free parameter in each fit, therefore the positions of the data points on the $x$-axis differ between fits, as is in particular evident for the spherical fit.}
\label{fig:example_event_fits}
\end{figure}

The resulting best fitting wavefront shapes are given in Fig.~\ref{fig:hyp_fit}, \ref{fig:con_fit} and \ref{fig:sph_fit} for a hyperbolic, conical and spherical wavefront, respectively.

\begin{figure}
	\centering
	\includegraphics[width=0.80\textwidth]{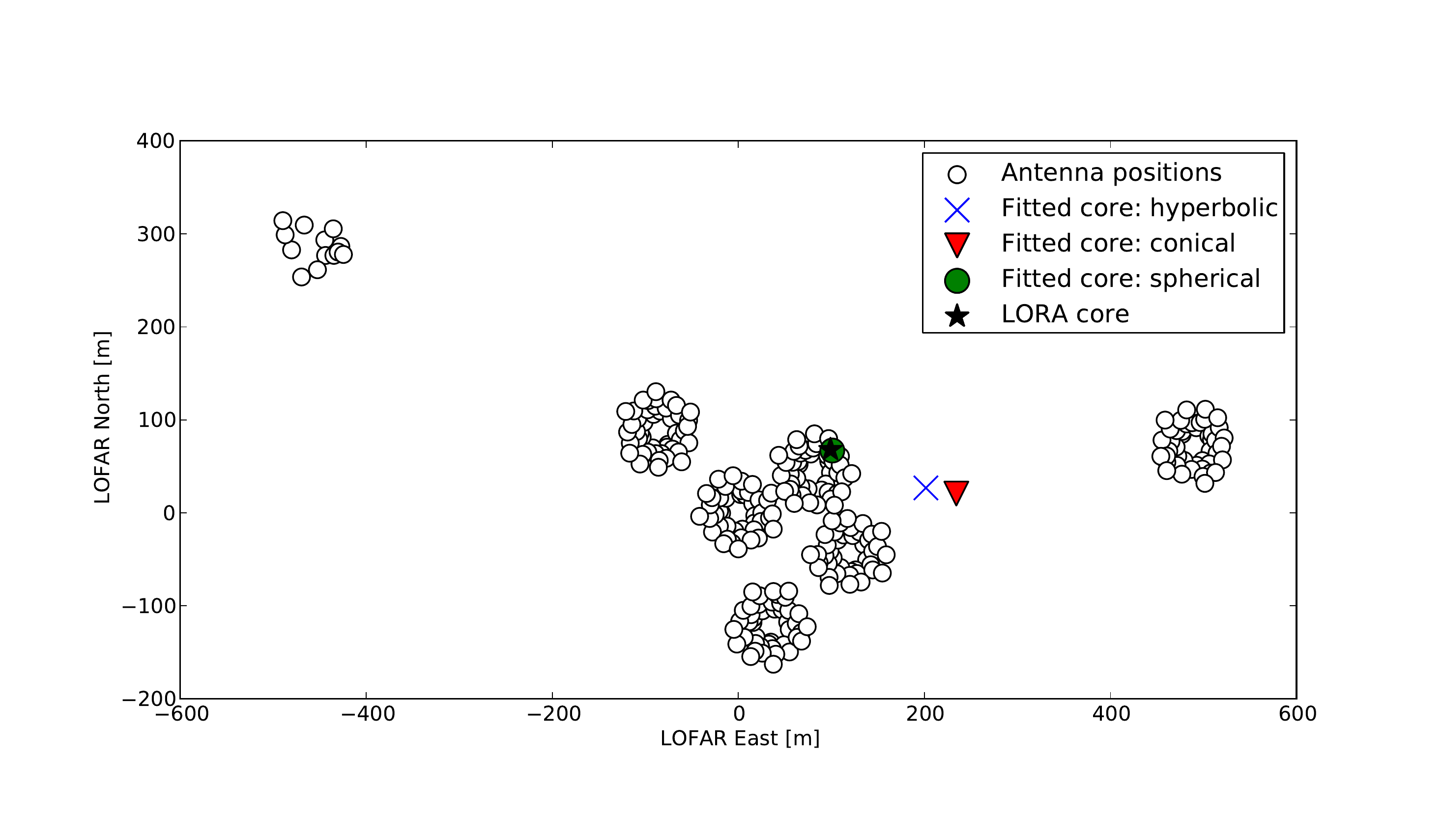}
	\caption{Fitted shower core positions for the shower in Fig.~\ref{fig:example_event_fits}, for the different wavefront shapes. Note that the core position determined by the LORA particle detector array is not reliable for this particular air shower since it is located at the edge (or even outside) of the instrumented area.}
\label{fig:fitted_core_positions}
\end{figure}

\begin{figure}
	\centering
	\includegraphics[width=0.80\textwidth]{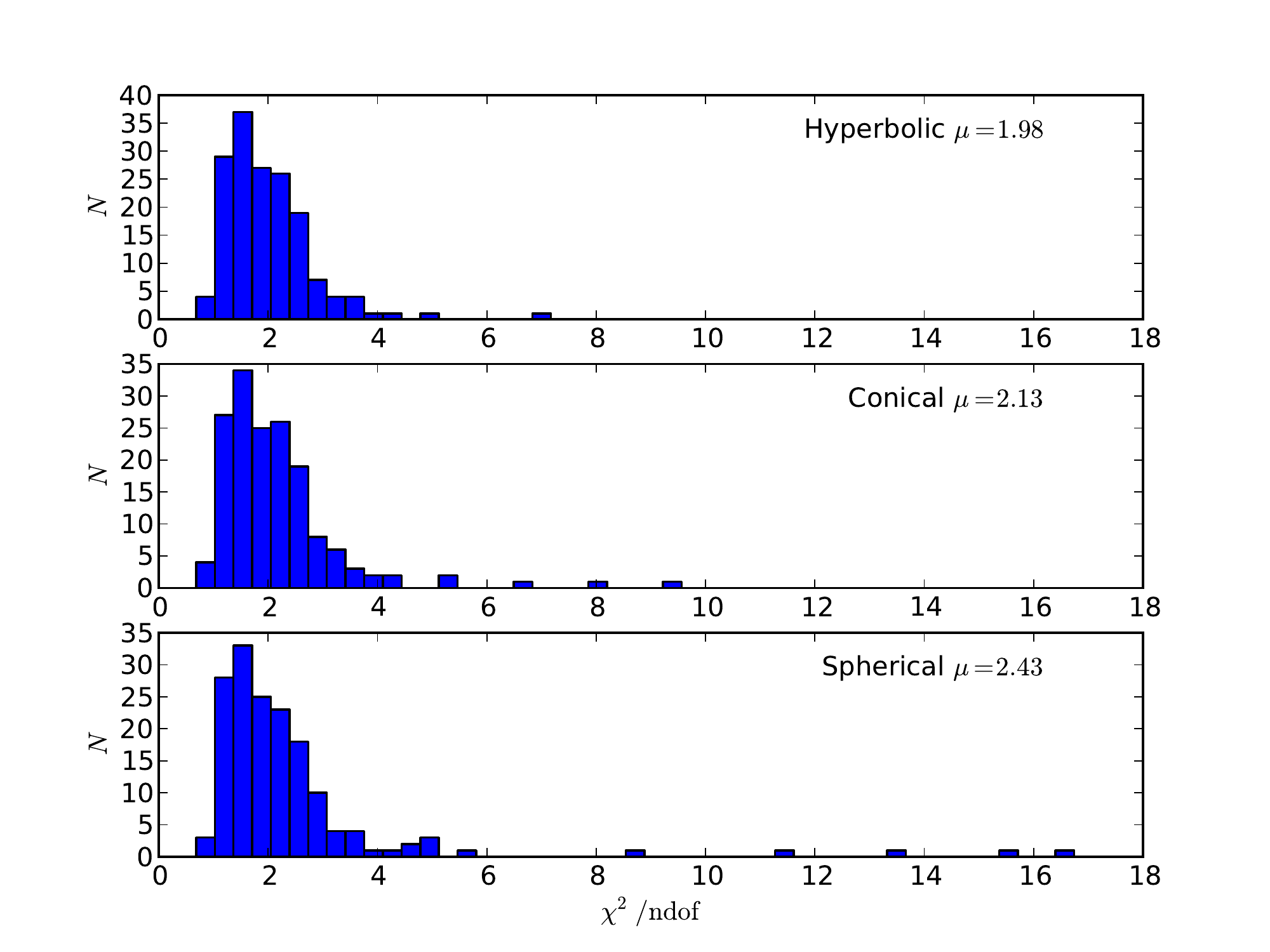}
	\caption{Fit quality for a hyperbolic (top), conical (middle) and spherical (bottom) wavefront shape.}
\label{fig:wf_chi2_distributions}
\end{figure}

The wavefront shape of this air shower is best fitted by a hyperbola due to significant curvature near the shower axis. The shower core position, left as free parameters in the fitting procedure, is significantly different for the three fits, as shown in Fig.~\ref{fig:fitted_core_positions}.
Fig.~\ref{fig:wf_chi2_distributions} shows the $\chi^2 / \mathrm{ndf}$ values obtained for all showers.
From these distributions it is not immediately evident which wavefront shape (if any) is favored. However, these distributions do not reflect the often significant differences in fit quality for a single shower. Furthermore, even if the wavefront shape were always hyperbolic one would still expect to see shapes that appear conical or spherical for individual showers depending on the shower geometry and the part of the shower front that is sampled by the detector.

In order to check which wavefront shape is favored by the overall dataset we perform a likelihood ratio test. The test statistic for the conical case is:
\begin{align}
D &= -2\,\frac{\ln(\mathrm{likelihood\, hyperbolic})}{\ln(\mathrm{likelihood\, conical})}\\
  &= \sum_{k}^{N} \chi_{\mathrm{con}}^2 - \chi_{\mathrm{hyp}}^2
\end{align}
where the sum $k$ is over all $N$ showers. For an appropriate choice of parameters the hyperbolic function can turn into either a conical or a spherical function. Thus, the solution space of the spherical and conical fit functions are subsets of the solution space of the hyperbolic fit. Therefore (if the fit converged correctly) the hyperbolic fit will always have a lower $\chi^2/\mathrm{ndf}$ value, even when the wavefront shape is intrinsically spherical or conical.

Under the null hypothesis that the wavefront shape is intrinsically conical (or spherical) the test statistic $D$ should follow a $\chi^{2}(N)$ distribution.
For large $N$, the $\chi^2(N)$ distribution approximates a Gaussian with mean $N$ and standard deviation $\sqrt{2 N} \ll D - N$.
From the data we obtain the value $D=6309$. The probability for this value to occur if the shape is conical is effectively zero, $p\ll 10^{-9}$, as the $D$-value is very far out of the distribution range.

There are two possible reasons for obtaining a higher value. Either the timing uncertainties are underestimated or the wavefront shape is generally not conical. Given the obtained reduced $\chi^{2}$ values of the hyperbolic fit, averaging to $1.98$, it is unlikely that the uncertainties are underestimated by more than a factor $\sim 1.5$. This is not enough by far to explain the measured value of the test statistic. Therefore we reject the null hypothesis that the wavefront shape is conical. Using the same procedure we also reject a spherical wavefront shape, with $D=16927$ and correspondingly an even (much) lower $p$-value.
Moreover, the lack of overall structure in the residuals of the hyperbolic fits (at our timing precision) argues against a more complicated wavefront shape. Therefore we conclude that the wavefront shape is hyperbolic. Furthermore, we do not see any evidence for a deviation from rotational symmetry (around the shower axis). So this is either not present, or is not resolvable with the current timing resolution.

\subsection{Direction reconstruction}
\begin{figure}
	\centering
	\begin{subfigure}[b]{0.80\textwidth}
	\includegraphics[width=\textwidth]{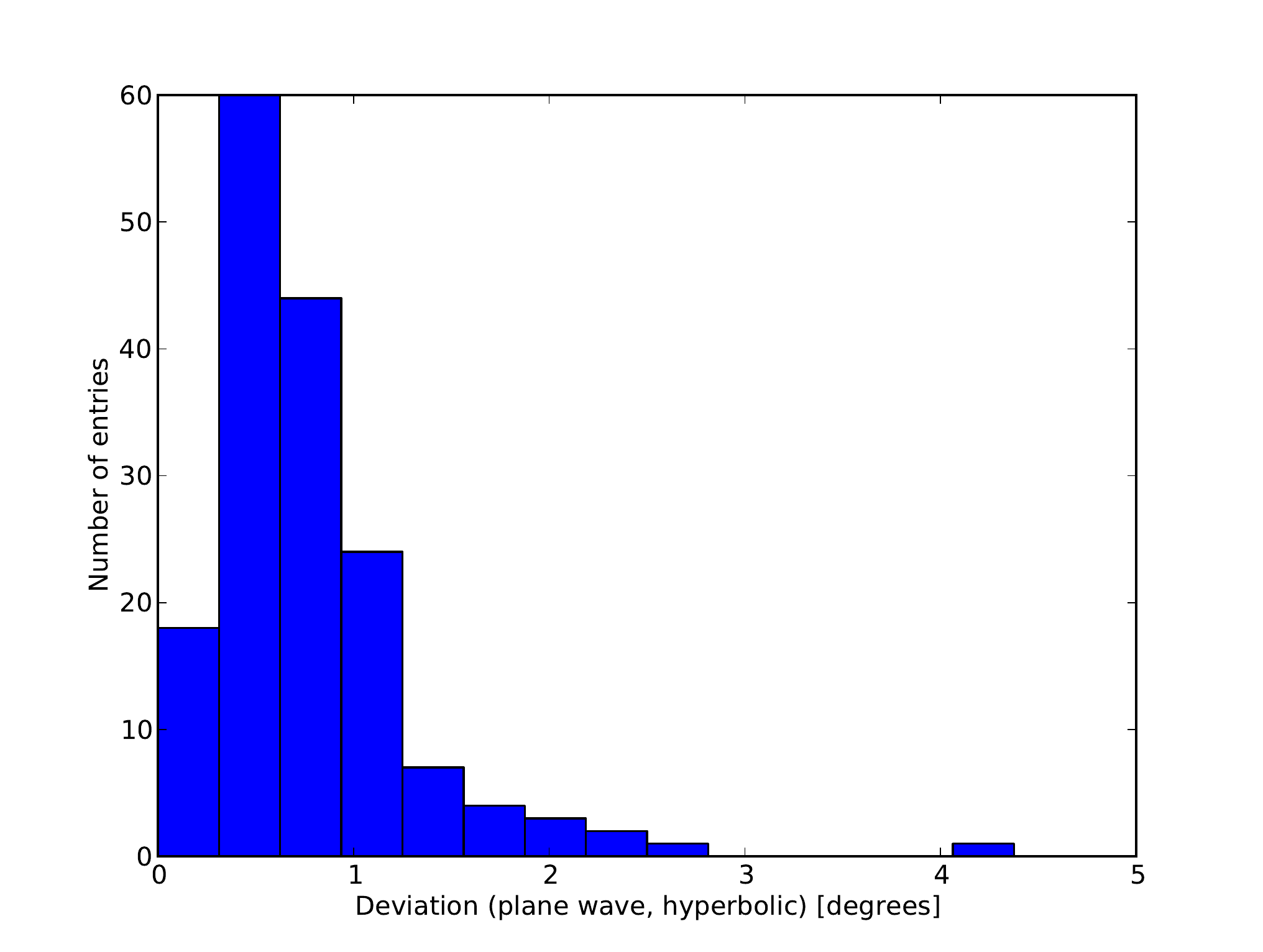}
	\end{subfigure}
	\begin{subfigure}[b]{0.80\textwidth}
	\includegraphics[width=\textwidth]{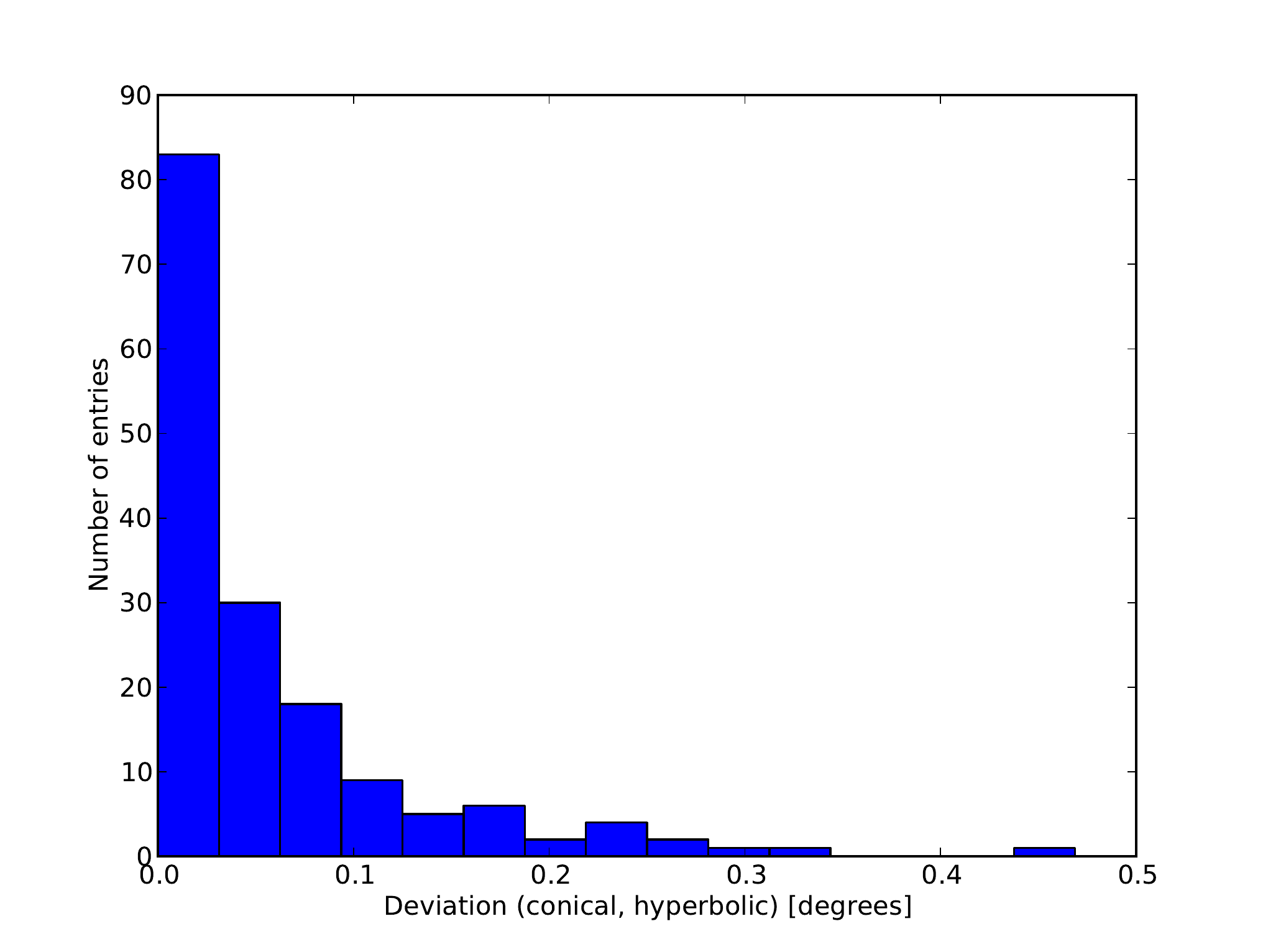}
	\end{subfigure}
	\caption[Direction precision]{Angular difference between reconstructed shower axis direction for three wavefront shape assumptions. Assuming a planar wavefront shape typically introduces an error in the direction of up to $\sim\unit[1]{^\circ}$, when the shape is in fact hyperbolic (top plot). The differences in reconstructed direction between a conical and hyperbolic wavefront shape are approximately a factor of ten smaller (bottom plot).}
\label{fig:directions}
\end{figure}

Comparing the reconstruction of the shower axis direction for the best fitting hyperbolic wavefront to the planar wavefront solution, in the top panel of Fig.~\ref{fig:directions}, we see that the latter deviates by up to $\sim\unit[1]{^\circ}$. Therefore, using a non-planar wavefront shape leads to an improvement in reconstruction precision of the air shower direction.

As can be seen from the bottom panel in Fig.~\ref{fig:directions} however, the exact shape of the non-planar wavefront is less important. The difference between the reconstructed direction using a conical or hyperbolic wavefront shape is typically less than $\unit[0.1]{^\circ}$. Since a conical fit contains one less free parameter this may be more practical in reconstruction software. However, a planar fit does not depend on the position of the shower axis or the exact shape of the wavefront and is thereby more robust and more suitable for standard reconstruction software when higher precision is not required.
\subsection{Correlations with air-shower parameters}
From \cite{Schroeder2011} it is expected that the shape of the radio wavefront depends on air-shower parameters and the distance to the shower maximum in particular. Since, for a shower with the same $X_\mathrm{max}$ the distance to shower maximum increases with increasing zenith angle ($\theta$), the shape of the radio wavefront is also expected to depend on the zenith angle.
This can be seen in Fig.\ \ref{fig:why_hyperbolic} where the radius of curvature of the inner part, its extent and the slope of the conical part are all expected to depend on the distance to the last emission point. This in turn would depend on $X_\mathrm{max}$.

\begin{figure}
\centering
\includegraphics[width=0.80\textwidth]{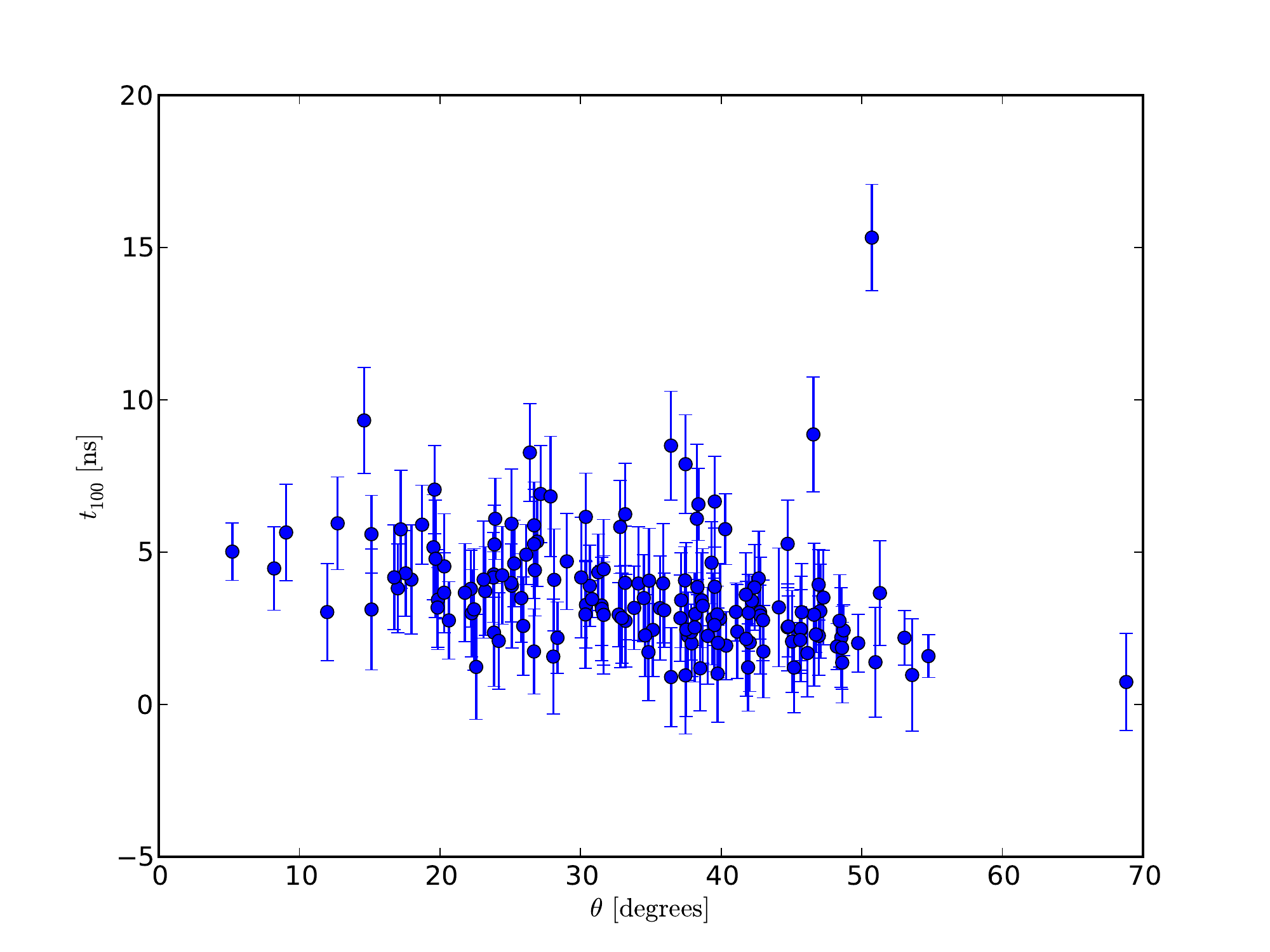}
\caption{Time lag of the radio wavefront at $r=\unit[100]{m}$, with respect to the arrival time of the emission along the shower axis ($r=0$), as a function of zenith angle. Arrival times are obtained by evaluating the best fitting hyperbolic fit at $r=\unit[100]{m}$. Uncertainties on $t_{100}$ represent one standard deviation of the scatter around the best fitting hyperbolic fit over the full range of $r$, which are typically a factor $\sim 1.5$ larger than the timing uncertainty for individual antennas.}
\label{fig:t_correlation}
\end{figure}

Similar to \cite{Schroeder2011}, we can take e.g.\ the time lag of the radio wavefront at $r=\unit[100]{m}$, with respect to the arrival time of the emission along the shower axis ($r=0$). It is not possible to use the hyperbola parameter $b$ (the slope of the asymptote) directly, as in some cases the asymptotic regime is (far) outside the data range.
Fig.\ \ref{fig:t_correlation} shows the time lag at $r=\unit[100]{m}$ as a function of zenith angle. We find a weak correlation with a Pearson correlation coefficient of $-0.32$. The probability of obtaining this value for uncorrelated data is $4\cdot 10^{-5}$. 

To give an order of magnitude of the angular deviation between the measured wavefront and the shower plane, we can use $t_{100}$ to get
\begin{equation}
\alpha = \frac{c\,t_{100}}{\unit[100]{m}},
\end{equation}
which is on average $\unit[0.11]{rad} = \unit[0.63]{^\circ}$.
As the hyperbola becomes steeper further out, we could also use $t_{250}$ instead (still inside the data range), which would give on average $\unit[0.94]{^\circ}$. These numbers agree qualitatively with the average deviation angle from a plane of $\unit[0.83]{^\circ}$ found by \cite{Schroeder2011}.
The small angle of less than one degree explains why precise timing is required in order to measure the wavefront shapes.

In practice however, it appears to be difficult to use wavefront timing by itself to determine (the distance to) $X_\mathrm{max}$. This is due to the strong interdependency of the shower axis position and the exact shape of the wavefront. While the wavefront shape remains hyperbolic when moving the shower axis location around, the curvature near the axis as well as the slope further out change. Therefore it is best to combine timing information with other information available on the shower. This information may come from the particle detectors, or from the radio data in the form of the intensity pattern at ground level. It has already been shown that the radio intensity pattern itself is highly sensitive to $X_\mathrm{max}$ \cite{ICRC_Buitink}. Combining this technique with timing information will improve the precision of these measurements.

\section{Conclusions}
We have shown that the wavefront of the radio emission in extensive air showers is measured to a high precision (better than $\unit[1]{ns}$ for each antenna) with the LOFAR radio telescope. The shape of the wavefront is best parametrized as a hyperboloid, curved near the shower axis and approximately
conical further out. A hyperbolic shape fits significantly better than the previously proposed spherical and conical shapes.

Reconstruction of the shower geometry using a hyperbolic wavefront yields a more precise determination of the the shower direction, and an independent measurement of the core position. Assuming the resulting reconstructed direction has no systematic bias, the angular resolution improves from $\sim 1\, ^\circ$ (planar wavefront) to $\sim 0.1\, ^\circ$ (hyperbolic). This assumption will be tested in a future simulation study.
This improvement will be of particular importance for radio $X_\mathrm{max}$ measurements for highly inclined showers where small deviations in arrival angle correspond to large differences in the slanted atmospheric depth.

The high antenna density and high timing resolution of LOFAR offer a unique opportunity for a detailed comparison with full Monte Carlo air shower simulations, including the arrival time measurements presented here. Furthermore, efforts to integrate timing information within the $X_\mathrm{max}$ measurement technique from \cite{ICRC_Buitink} are currently ongoing.

\section{Acknowledgements}
The authors thank Eric Cator for a useful discussion on statistical tests, and thank the anonymous referee for constructive comments. 
The LOFAR Key Science Project Cosmic Rays very much acknowledges the scientific and technical support from ASTRON.

Furthermore, we acknowledge financial support from the Netherlands Research School for Astronomy (NOVA), the Samenwerkingsverband Noord-Nederland (SNN) and the Foundation for Fundamental Research on Matter (FOM) as well as support from the Netherlands Organization for Scientific Research (NWO), VENI grant 639-041-130. We acknowledge funding from an Advanced Grant of the European Research Council under the European Union's Seventh Framework Program (FP/2007-2013) / ERC Grant Agreement n.\ 227610.
Chiara Ferrari acknowledges financial support by the {\it ``Agence Nationale de la Recherche''} through grant ANR-09-JCJC-0001-01.

LOFAR, the Low Frequency Array designed and constructed by ASTRON, has facilities in several countries, that are owned by various parties (each with their own funding sources), and that are collectively operated by the International LOFAR Telescope (ILT) foundation under a joint scientific policy.

\bibliographystyle{elsarticle-num}
\bibliography{wavefront_curvature}

\end{document}